# Dissolution-Precipitation Growth of Uniform and Clean Two Dimensional Transition Metal Dichalcogenides

Zhengyang Cai[1,†], Yongjue Lai[1,†], Shilong Zhao[1], Rongjie Zhang[1], Junyang Tan[1], Simin Feng[1], Jingyun Zou[1], Lei Tang[1], Junhao Lin[3], Bilu Liu[1,*], Hui-Ming Cheng[1,2,*]

[1]Shenzhen Geim Graphene Center, Tsinghua−Berkeley Shenzhen Institute and Tsinghua Shenzhen International Graduate School, Tsinghua University, Shenzhen 518055, China

[2]Shenyang National Laboratory for Materials Sciences, Institute of Metal Research, Chinese Academy of Sciences, Shenyang 110016, China

[3]Department of Physics, Southern University of Science and Technology, Shenzhen 518055, China

[†]These authors contribute equally.

[*]Corresponding authors:

bilu.liu@sz.tsinghua.edu.cn (BL); hmcheng@sz.tsinghua.edu.cn (HMC)

**ABSTRACT**

Two-dimensional (2D) transition metal dichalcogenides (TMDCs) have attracted much interest and shown promise in many applications. However, it is challenging to obtain uniform TMDCs with clean surfaces, because of the difficulties in controlling the way the reactants are supplied to the reaction in the current chemical vapor deposition (CVD) growth process. Here, we report a new growth approach called "dissolution-precipitation" (DP) growth, where the metal sources are sealed inside glass substrates to control their feeding to the reaction. Noteworthy, the diffusion of metal source inside glass to its surface provides a uniform metal source on the glass surface, and restricts the TMDC growth to only a surface reaction while eliminates unwanted gas-phase reaction. This feature gives rise to highly-uniform monolayer TMDCs with a clean surface on centimeter-scale substrates. The DP growth works well for a large variety of TMDCs and their alloys, providing a solid foundation for the controlled growth of clean TMDCs by the fine control of the metal source.

**INTRODUCTION**

Two dimensional (2D) transition metal dichalcogenides (TMDCs) have attracted increasing attention due to their atomically-thin body, excellent electronic and optoelectronic properties, and abundant material choices.[1-4] Chemical vapor deposition (CVD) is an important method of preparing TMDCs and great success has been achieved in growing large single crystals as well as continuous films.[5-9] Currently, one bottleneck in the CVD growth of TMDCs is that it is difficult to prepare uniform, large-area and ultraclean monolayer TMDCs, because the metal sources can hardly be precisely controlled using current feeding methods.[10-12] In a typical TMDC growth process, solid sources like $MoO_3$ and S powders are used. First, the feed amount of Mo is location dependent, which means that the Mo concentration is different at different positions on the substrate, causing a non-uniform $MoS_2$ distribution.[10, 13] Second, the $MoO_3$ and S feeds share the same diffusion path so that there may be gas-phase reactions in addition to the on-substrate reaction, causing by-product deposited on surface of as-grown $MoS_2$. There has been much effort to solve these problems, such as using liquid- or gas-phase Mo and S sources, and pre-deposition of the Mo source.[14-16] Nevertheless, it is still difficult to grow uniform and ultraclean TMDCs over large areas.[17, 18]

To tackle these issues, we may learn the lesson from graphene growth.[19-21] In typical graphene growth by CVD, the carbon source is dissolved in the bulk or sub-surface of catalytic substrates like nickel or copper, followed by precipitation at the substrate surface to grow uniform graphene over large areas.[22, 23] This mechanism also works well for other 2D materials. For example, Shi et al. have used a molten $Fe_{82}B_{18}$ alloy which supplies boron source and dissociates nitrogen in the carrier gas for the growth of multilayer boron nitride.[24] Li et al. have recently reported that $MoS_2$ ribbons can be grown by forming Na-Mo-O droplets through a vapor-liquid-solid growth mechanism.[25, 26] Therefore, analogous to the mechanism for growing graphene, it is possible to grow uniform and clean 2D TMDCs, if one can "dissolve" the metal source into the growth substrate to control its feed.[27]

In this work, we report a "dissolution-precipitation" (DP) growth method that achieves the dissolution of the required metal source into the growth substrate and succeeds in growing uniform and clean monolayer TMDCs. In this method, the metal source is embedded between two pieces of glass, and gradually diffuses out to the surface of the upper glass during growth. In this way, we have (i) achieved a uniform feed of the metal source and (ii) restricted the reaction to only the surface of the top glass while eliminated any unwanted gas-phase reactions because the metal and chalcogen sources do not share the same diffusion path.[23, 28] As a result of these two features, highly uniform monolayer TMDCs with a clean surface have been grown on a centimeter-scale molten glass surface. The method has been used for many different TMDCs and their alloys, such as $MoSe_2$, $WS_2$, $MoTe_2$, $Mo_xW_{1-x}S_2$, and V doped $MoS_2$, showing good universality.

**RESULTS AND DISCUSSIONS**

A scheme of the DP growth of TMDCs is illustrated in **Figure 1a**. A metal source

(e.g., $Na_2MoO_4$, $Na_2WO_4$, $NaVO_3$) was embedded between two pieces of glass (the thickness of the bottom is 2 mm and the top one is 0.15 mm) to make a glass/metal source/glass sandwich structure (SG-M), which was then heated and fused together. This sandwich was used as a substrate and the metal source for the growth of TMDCs. The chalcogen source (S, Se, or Te powder) was placed at the upper steam in a horizontal tube furnace and the growth was conducted at 700-800 °C depending on the material used. With increasing temperature, the metal source melted and diffused through the molten glass to its surface[28-30], which is called the "dissolution" process. An AFM image of a typical surface of the top glass (**Figure 1a**) shows a protrusion with a lateral size of ~2 μm and a height of ~20 nm, which serves as the metal source for subsequent TMDC growth. From the volume of a protrusion, we estimate that the amount of Mo precursor in each protrusion is around $1.8 \times 10^{-13}$ g. Such a low concentration of metal source in the feed has been effective in many CVD process.[28, 31, 32] After diffusing to the upper surface, the metal sources react with the chalcogen to grow TMDCs on the molten glass surface, which is called the "precipitation" process. The details of the growth mechanism are discussed in **Figures S1-7**. In the nucleation stage, the liquid-phase metal source uniformly diffuses to surface (**Figure 1b**), guaranteeing a uniform supply. In contrast, in the traditional CVD growth of TMDCs where the metal precursor is a solid powder which sublimates in a hard-to-control manner, the metal source concentration depends on the distance from the solid source (**Figures 1c and S8**).[13] In addition, during DP growth, the metal source and chalcogen vapor meet only on the substrate surface, so only here does the reaction happen (**Figure 1b**). This is distinct from traditional CVD where the metal source and chalcogen vapor first meet in the gas phase and both gas-phase and surface reactions occur, resulting in the non-uniform nucleation and growth of TMDCs, as well as deposition of gas phase products on the TMDC surface (**Figure 1c**). Therefore, this DP method grows TDMCs that are both uniform and clean compared to the traditional CVD method.

To investigate the effects of the metal source feeding process in the DP method, we first studied the uniformity of as-grown $MoS_2$ over a 2.5 cm × 1.0 cm substrate. A movie (**Movie S1**) shows the distribution of flakes on the whole substrate, indicating a highly uniform distribution. We analyzed the nuclei density (**Figure 2a**), average area (**Figure 2b**) and perimeter (**Figure S10**) of each image extracted from the movie, and found that all show narrow distributions. The average nucleation density is 1080 flakes/$mm^2$, the average perimeter of individual $MoS_2$ flake is 47 μm, and the average area is 92 $μm^2$. These values are highly uniform for 140 images over 2.5 cm × 1.0 cm area (**Figure S11**). The above results have demonstrated that $MoS_2$ grown by the DP method is highly uniform over a centimeter-scale substrate.

We then studied the quality and uniformity of the DP-grown TMDC flakes. First, we randomly chose 50 flakes from a molten glass substrate and analyzed their $E_{2g}$ and $A_{1g}$ Raman peaks (**Figure 2c**). All the $MoS_2$ flakes had an $E_{2g}$ peak in the range 384.0 to 386.5 $cm^{-1}$, and an $A_{1g}$ peak in the range 402 to 404 $cm^{-1}$, with the frequency differences between $E_{2g}$ and $A_{1g}$ peaks in a narrow range of 17.6-18.6 $cm^{-1}$. Second, we measured photoluminescence (PL) spectra of the same 50 flakes and found that most of A exciton peaks of MoS$_2$ were distributed from 684 to 689 nm (**Figure 2d**). Both results indicate the uniformity of the DP-grown TMDCs. Third, we investigated an individual MoS$_2$ flake. The PL intensity map at 686.8 nm (**Figure 2f**) and Raman intensity map in the A$_{1g}$ mode at 403.1 cm$^{-1}$ (**Figure 2g**) for the area shown in **Figure 2e** show a quite uniform intensity over the whole flake. We noticed that for an individual MoS$_2$ flake on molten glass, (i) the Raman spectrum showed an E$_{2g}$ at 385.2 cm$^{-1}$ and an A$_{1g}$ at 402.8 cm$^{-1}$ with a difference of 17.6 cm$^{-1}$, smaller than that of monolayer MoS$_2$ grown on a SiO$_2$/Si substrate, (ii) the A exciton peak located at 686.8 nm showed a red shift compared with MoS$_2$ grown on a SiO$_2$/Si substrate. These changes may be ascribed to a strain effect and dielectric screening between the monolayer MoS$_2$ and molten glass,[33-35] since after the flake was transferred onto a SiO$_2$/Si substrate, the Raman peaks and A exciton of DP-grown MoS$_2$ are similar to those of exfoliated monolayer ones (**Figures S12 and S13**). To sum up, the above results indicate that the MoS$_2$ grown using the DP method is a uniform monolayer.

We then investigated the surface cleanness of the DP-grown MoS$_2$ flakes. First, AFM was used to characterize the surface flatness. For MoS$_2$ flakes grown using traditional CVD with solid MoO$_3$ powder as the metal source, many small particles were observed at the edges as well as on the plane (**Figure 3a**). This feature is reported in the literature using similar methods.[5, 13, 36, 37] In sharp contrast in **Figure 3b**, the MoS$_2$ flakes grown using the DP method exhibit a clean surface except for a protrusion under each flake which is the Mo precursor confirmed by the AFM images of transferred MoS$_2$ in **Figure 3c.** Second, we exposed the surface of the substrate on which the MoS$_2$ flakes had grown to TiCl$_4$ vapor in humid air. Since TiCl$_4$ is easily hydrolyzed to form TiO$_2$ particles, which will be selectively absorbed by contaminated area of MoS$_2$.[38] As shown in **Figures 3d and e,** the flakes grown by traditional CVD had many TiO$_2$ particles while the DP-grown flakes showed only a few. These observations also show that the DP-grown MoS$_2$ are much cleaner than traditional CVD-grown ones. Third, we checked the interlayer coupling of two monolayer MoS$_2$ stacked structures. Generally, two stacked MoS$_2$ layers with a clean interface will have a strong interlayer coupling characterized by suppressed monolayer PL emissions and the appearance of interlayer optical transitions.[39, 40] We observed both a clear suppression of the A exciton emissions at ~660 nm and the emergence of an interlayer emission peak at around 740 nm in the stacked bilayer MoS$_2$ grown by the DP method (**Figure 3f**), which indicates a clean surface of the monolayers. This phenomenon is in striking contrast to what was observed when stacking layers grown by traditional CVD (**Figure S14**), where no interlayer coupling was observed. The clean surface grown by the DP method originated from the reaction between the surface-limited diffusion of the Mo source and the S vapor which is supplied in a different gas-phase path, therefore secondary nucleation process is prohibited.

We also characterized the quality and electrical performance of the DP-grown MoS$_2$. Scanning-transmission electron microscopy (STEM) images show the DP-grown MoS$_2$ maintain perfect hexagonal lattice without apparent defects. The corresponding fast Fourier transform (FFT) pattern confirms the 2H phase of MoS$_2$ (**Figures 3g and 3h**). The XPS results for the DP-grown MoS$_2$ show the typical binding energies of Mo

3d$_{5/2}$ (229.5 eV), Mo 3d$_{3/2}$ (232.7 eV), S 2p$_{3/2}$ (162.3 eV), and S 2p$_{1/2}$ (163.6 eV), with a Mo : S atomic ratio of 1:1.97 (**Figure S15**). We also fabricated several field-effect transistors using the DP-grown MoS$_2$ (**Figures 3i and S16-17**), which showed a decent carrier mobility in range of 7.5-21.5 cm$^2$V$^{-1}$s$^{-1}$ and an on/off ratio in range of 10$^6$-10$^8$. These results confirm the high quality of the DP-grown MoS$_2$ flakes which are comparable to other growth techniques.

We found that DP is a universal method to grow various TMDCs besides MoS$_2$, as well as to grow TMDC alloys and doped TMDCs. As shown in **Figure 4a**, DP-grown MoSe$_2$ flakes have a typical triangular shape with a size ranging from 10 to 25 μm. **Figure 4b** shows the two characteristic peaks (A$_{1g}$ and E$_{2g}$) of MoSe$_2$ at 239.6 cm$^{-1}$ and 287.2 cm$^{-1}$, respectively. There is no peak at around 350 cm$^{-1}$, indicating that the as-grown flakes are monolayer MoSe$_2$. The PL spectrum (**Figure 4c**) shows a direct bandgap peak at 789.8 nm.[40, 41] This method also works well for MoTe$_2$, which is not easy to grow by the traditional CVD method. **Figures 4d and e** show the needle-like shape of 1T' phase MoTe$_2$ and its typical Raman spectrum, which agrees well with that of mechanically exfoliated 1T' MoTe$_2$.[42] We have also extended this DP method to grow monolayer WS$_2$ (**Figure 4f**). An E$_{2g}$ peak of WS$_2$ at 358.6 cm$^{-1}$ and an A$_{1g}$ peak at 418.7 cm$^{-1}$ are observed (**Figure 4g**). The frequency difference between these two modes is about 60.1 cm$^{-1}$ and the PL peak position is located at 621.3 nm (**Figure 4h**), good matches with the values for monolayer WS$_2$.[43] The above results demonstrate the versatility of the DP method in growing various TMDCs.

Furthermore, the DP growth method can be used to grow Mo$_x$W$_{1-x}$S$_2$ alloy, and V-

doped MoS$_2$. An optical image of Mo$_x$W$_{1-x}$S$_2$ is shown in **Figure 4i**. The Raman spectrum contains characteristic peaks of both WS$_2$ and MoS$_2$, where the E$_{2g}$ peak and A$_{1g}$ peak of MoS$_2$ are located at 385.8 cm$^{-1}$ and 404.8 cm$^{-1}$, respectively, and those of WS$_2$ are located at 357.4 cm$^{-1}$ and 418.3 cm$^{-1}$, respectively (**Figure 4j**). The PL peak of the Mo$_x$W$_{1-x}$S$_2$ alloy is located at 639.4 nm as shown in **Figure 4k**, which is between the wavelengths of pristine MoS$_2$ and WS$_2$.[44] Recent study shows that a V-doped WSe$_2$ monolayer is a room temperature ferromagnetic semiconductor,[45] and the DP method can be used to grow such flakes. As shown in **Figure S18**, both low and high concentration V-doped MoS$_2$ monolayers were grown. The effective doping of V was further verified by the XPS results shown in **Figure S19**. Taken together, these results have demonstrated that the DP method is a general approach to grow a family of TMDCs.

**CONCLUSION**

We have developed a DP method for TMDC growth. In this method, the metal source is provided by diffusion through a thin molten glass substrate, leading to a uniform distribution of metal precursor and restricting the growth to only a surface reaction. As a result, highly uniform and monolayer TMDCs with clean surfaces have been grown on centimeter-scale glass substrates. We have also extended the method to the growth of other TMDCs to demonstrate its universality. These results highlight a new general approach for the growth of uniform and clean 2D materials for widespread applications.

**METHODS**

**Embedding the metal precursor inside a glass substrate.** First, a $Na_2MoO_4$ (0.94 mg) solution (4 μL, 1 mol/L in DI water) was dropped onto a 2-mm-thick soda lime glass slide with a size of 1.0 cm×1.0 cm or 1.0 cm×2.5 cm, and then dried in an oven. Then, a thinner (0.15-mm- thick) soda lime glass slide of the same size was then placed on top of the above thick glass and the two were heated in a muffle furnace at 10 °C/min to 660 °C where they were sintered for 30 min to join them with the precursors remaining in between. For other studies the $Na_2MoO_4$ was replaced with $Na_2WO_4$ or a mixture of $Na_2MoO_4$ and $Na_2WO_4$ or a mixture of $Na_2MoO_4$ and $NaVO_3$. The above fused glass sandwich was used as both the growth substrate and metal source for the DP growth. Note that the fused glass with the metal precursor inside is named SG-M, where M is Mo or W or a mixture.

**DP growth of $MoS_2$.** A horizontal tube furnace was used for the DP growth of $MoS_2$. S powder (150 mg, 99.5%, Sigma-Aldrich) was loaded upstream where the temperature was 150 °C and the SG-Mo was placed in the center of the furnace, serving as Mo precursor and growth substrate simultaneously. Before growth, the tube was pumped to 0.05 Torr and refilled with Ar to ambient pressure which was repeated three times to eliminate residual oxygen and water. During the growth, the temperature was increased to 730-750 °C at a rate of 50 °C/min and the growth lasted for 10-20 min. Ar was used as the carrier gas with a flow rate of 80 sccm at 1.2-2.0 Torr. After growth, the furnace was cooled to 200 °C under 80 sccm of Ar flow.

**DP growth of other TMDCs and alloys.** For the DP growth of MoSe$_2$, MoTe$_2$, WS$_2$, Mo$_x$W$_{1-x}$S$_2$ and V-doped MoS$_2$, we followed a similar growth procedure to that used for MoS$_2$ but with slight modifications. **(i) MoSe$_2$ growth:** Se powder (200 mg, 99.5%, Sigma-Aldrich) was loaded in the furnace position where the temperature was 280 °C. The growth temperature was 750-800 °C. 80 sccm Ar and 8 sccm H$_2$ were used as the carrier gas at a low pressure of 1.2-2.0 Torr. **(ii) MoTe$_2$ growth:** Te powder (200 mg, 99.5%, Sigma-Aldrich) was loaded upstream about 11 cm away from the center of the furnace. The growth temperature was 730-750 °C. 120 sccm Ar and 20 sccm H$_2$ were used as the carrier gas at ambient pressure. **(iii) WS$_2$ growth:** S powder (150 mg, 99.5%, Sigma-Aldrich) was loaded upstream where the temperature was 190 °C, and SG-W was placed in the center of the furnace. The growth temperature was 730-780 °C. 80 sccm Ar and 4 sccm H$_2$ were used as the carrier gas at ambient pressure. **(iv) Mo$_x$W$_{1-x}$S$_2$ growth:** S powder (150 mg, 99.5%, Sigma-Aldrich) was loaded upstream where the temperature was 150 °C. SG-Mo-W (the ratio of Na$_2$MoO$_4$ : Na$_2$WO$_4$ was 1:15) was placed in the center of the furnace. The growth temperature was 750-780 °C. 80 sccm Ar and 4 sccm H$_2$ were used as the carrier gas at a low pressure of 1.2-2.0 Torr. **(v) V-doped MoS$_2$ growth:** S powder (120 mg, 99.5%, Sigma-Aldrich) was loaded upstream where the temperature was 150 °C. SG-Mo-V (for low concentration V doping, the ratio of Na$_2$MoO$_4$ : NaVO$_4$ was 1:1, for high concentration V doping, the ratio of Na$_2$MoO$_4$ : NaVO$_4$ was 1:5) was placed in the center of the furnace. The growth temperature was 730-750 °C. 120 sccm Ar and 8 sccm H$_2$ were used as the carrier gas at a low pressure of 1.2-2.0 Torr.

**Growth of MoS$_2$ by traditional CVD**. A horizontal tube furnace was used for the growth of MoS$_2$. S powder (100 mg, 99.5%, Sigma-Aldrich) was loaded upstream at 220 °C. MoO$_3$ powder (10 mg, 99.5%, Sigma-Aldrich) was placed in the center of the furnace. SiO$_2$/Si (Si substrate with a 300 nm thick thermally-grown oxide) was used as growth substrate and placed on top of MoO$_3$ powder with the oxide layer facing down. The temperature was increased to the growth temperature of 700 °C at a rate of 50 °C min$^{-1}$. During growth, Ar was used as the carrier gas with a flow rate of 80 sccm at ambient pressure, and the growth lasted for 10 min.

**Transfer of TMDCs.** A polyethylene terephthalate (PET) film was placed on the obtained TMDC/glass and its temperature was increased to 65 °C. When the PET had softened, the TMDC became attached to PET to form PET/TMDC/glass. The removal of PET/TMDC from the glass substrate is through a very slow mechanical peeling followed by placing on a target substrate and heated to above 90 °C to cause the PET/TMDC to stick tightly to the target substrate. The PET/TMDC/target substrate was then immersed in dichloromethane to get rid of the PET, leaving the TMDC on the target substrate. Here, the target substrates could be either SiO$_2$/Si or TEM grids.

**Material Characterization.** The morphology and surface of the samples were examined by an optical microscope (Carl Zeiss Microscopy, Germany), SEM (5 kV, Hitachi SU8010, Japan) and AFM (Cypher ES, Asylum Research, USA). Raman and PL spectra were collected using 532 nm laser excitation with a beam size of ~1 μm (Horiba LabRAB HR Evolution Japan). Chemical elemental analyses of the samples were conducted by XPS (monochromatic Al Kα X-rays, 1486.6 eV, PHI VersaProbe II,

Japan). HAADF-STEM images were taken by an aberration-corrected TEM (FEI Titan Cube Themis G2300 with a field emission gun at 60 kV, USA), with a resolution of 0.8 Å. The acquisition parameters were set as below, i.e., probe size of 9, condenser lens aperture of 50 μm, and camera length of 145 mm.

**Device fabrication and measurements**. FET devices were fabricated using a laser writing system (Aresis Dell, ZKS). In brief, a drop of AZ5214 photoresist (PR) was spin-coated onto the $SiO_2$/Si substrate with the samples attached (2000 rpm for 1 min), and baked at 125 °C for 1 min. Photo-lithography was then conducted using PR as a positive resist. This was followed by successive develop, metal deposition, and lift-off to fabricate the FET devices. The metal electrodes were made of 5 nm Cr and 50 nm Au, which were deposited using e-beam evaporation. The device measurements were performed in a vacuum probe station ($10^{-5}$ mBar, Lakeshore TTPX, USA).

## SUPPLEMENTARY DATA

Supplementary data are available at NSR online.

## FUNDING

This work was supported by the NSFC of China (Nos. 51722206, 51991340, 51991343, 11974156, and 51920105002), the Youth 1000-Talent Program of China, the National Key R&D Program (2018YFA0307200), the Guangdong Innovative and Entrepreneurial Research Team Program (No. 2017ZT07C341), Guandong International Science Collaboration Project (Grant No. 2019A050510001), the Bureau of Industry and Information Technology of Shenzhen for the "2017 Graphene Manufacturing Innovation Center Project" (No. 201901171523), the Development and Reform Commission of Shenzhen Municipality for the development of the "Low-Dimensional Materials and Devices" discipline, and also the assistance of SUSTech Core Research Facilities, especially technical support from Pico-Centre that receives support from Presidential fund and Development and Reform Commission of Shenzhen Municipality.

**AUTHOR CONTRIBUTIONS**

B.L. and H.-M.C. supervised the project and directed the research. Z.C., B.L., and H.-M.C. conceived the idea. Z.C., Y.L., J.Z., and S.F., grew the materials and performed Raman, XPS, AFM, and SEM characterization. J.T. and J.L. performed TEM characterization. S.Z. and Y.L. performed optical measurements. R.Z. performed electrical measurements. Z.C., Y.L., H.-M.C., and B.L. interpreted the results and wrote the manuscript with feedbacks from the other authors.

**REFERENCE**

1. Chhowalla M, Shin HS, Eda G, et al.; The Chemistry of Two-Dimensional Layered Transition Metal Dichalcogenide Nanosheets. *Nat. Chem.* 2013;**5**(4):263-275.
2. Cao Y, Fatemi V, Fang S, et al.; Unconventional Superconductivity in Magic-Angle Graphene Superlattices. *Nature* 2018;**556**(7699):43.
3. Liu Y, Guo J, Zhu E, et al.; Approaching the Schottky–Mott Limit in van der Waals Metal–Semiconductor Junctions. *Nature* 2018;**557**(7707):696.
4. Zhang C, Tan J, Pan Y, et al.; Mass Production of Two-Dimensional Materials by Intermediate-Assisted Grinding Exfoliation. *National Science Review* 2020:**7**:324-332.
5. Cai Z, Shen T, Zhu Q, et al.; Dual-Additive Assisted Chemical Vapor Deposition for the Growth of Mn-Doped 2D $MoS_2$ with Tunable Electronic Properties. *Small* 2019; 1903181.
6. Yang P, Zou X, Zhang Z, et al.; Batch Production of 6-inch Uniform Monolayer Molybdenum Disulfide Catalyzed by Sodium in Glass. *Nat. Commun.* 2018;**9**(1):979.
7. Cai Z, Liu B, Zou X, et al.; Chemical Vapor Deposition Growth and Applications of Two-Dimensional Materials and Their Heterostructures. *Chem. Rev.* 2018;118(13): 6091-6133.
8. Gao Y, Liu ZB, Sun DM, et al.; Large-Area Synthesis of High-Quality and Uniform Monolayer $WS_2$ on Reusable Au Foils. *Nat. Commun.* 2015;**6**:8569.
9. Zhang Z, Chen P, Yang X, et al.; Ultrafast Growth of Large Single Crystals of Monolayer $WS_2$ and $WSe_2$. *National Science Review* 2020;nwz223, https://doi.org/10.1093/nsr/nwz223
10. Lee YH; Synthesis of Large-Area $MoS_2$ Atomic Layers with Chemical Vapor Deposition. *Adv. Mater.* 2012;**24**:2320-2325.
11. Liu B, Fathi M, Chen L, et al.; Chemical Vapor Deposition Growth of Monolayer $WSe_2$ with Tunable Device Characteristics and Growth Mechanism Study. *ACS Nano* 2015;**9**(6):6119-6127.
12. Zhou D, Shu H, Hu C, et al.; Unveiling the Growth Mechanism of $MoS_2$ with Chemical Vapor Deposition: From Two-Dimensional Planar Nucleation to Self-Seeding Nucleation. *Crystal Growth & Design* 2018;**18**(2):1012-1019.
13. Wang SS, Rong YM, Fan Y, et al.; Shape Evolution of Monolayer $MoS_2$ Crystals Grown by Chemical Vapor Deposition. *Chem. Mater.* 2014;**26**(22):6371-6379.
14. Lee JS, Choi SH, Yun SJ, et al.; Wafer-Scale Single-Crystal Hexagonal Boron Nitride Film via Self-Collimated Grain Formation. *Science* 2018;**362**(6416):817-821.
15. Eichfeld SM, Hossain L, Lin Y-C, et al.; Highly Scalable, Atomically Thin $WSe_2$ Grown via Metal-Organic Chemical Vapor Deposition. *ACS Nano* 2015;**9**(2):2080-2087.
16. Lim YR, Song W, Han JK, et al.; Wafer-Scale, Homogeneous $MoS_2$ Layers on Plastic Substrates for Flexible Visible-Light Photodetectors. *Adv. Mater.* 2016;**28**(25):5025-5030.
17. Jia K, Zhang J, Lin L, et al.; Copper-Containing Carbon Feedstock for Growing Superclean Graphene. *J. Am. Chem. Soc.* 2019;**141**(19):7670-7674.
18. Lin L, Zhang J, Su H, et al.; Towards Super-Clean Graphene. *Nat. Commun.*

2019;**10**(1):1912.

19. Wu TR, Zhang XF, Yuan QH, et al.; Fast Growth of Inch-Sized Single-Crystalline Graphene from a Controlled Single Nucleus on Cu-Ni Alloys. *Nat. Mater.* 2016;**15**(1):43-48.

20. Lee J-H, Lee EK, Joo W-J, et al.; Wafer-Scale Growth of Single-Crystal Monolayer Graphene on Reusable Hydrogen-Terminated Germanium. *Science* 2014;**344**(6181):286-289.

21. Xu X, Zhang Z, Dong J, et al.; Ultrafast Epitaxial Growth of Metre-Sized Single-Crystal Graphene on Industrial Cu Foil. *Science bulletin* 2017;**62**(15):1074-1080.

22. Reina A, Jia X, Ho J, et al.; Large Area, Few-Layer Graphene Films on Arbitrary Substrates by Chemical Vapor Deposition. *Nano Lett.* 2008;**9**(1):30-35.

23. Li X, Cai W, Colombo L, et al.; Evolution of Graphene Growth on Ni and Cu by Carbon Isotope Labeling. *Nano Lett.* 2009;**9**(12):4268-4272.

24. Shi Z, Wang X, Li Q, et al.; Vapor–Liquid–Solid Growth of Large-Area Multilayer Hexagonal Boron Nitride on Dielectric Substrates. *Nat. Commun.* 2020;**11**(1):1-8.

25. Li S, Lin YC, Zhao W, et al.; Vapour-Liquid-Solid Growth of Monolayer $MoS_2$ Nanoribbons. *Nat. Mater.* 2018;**17**(6):535-542.

26. Li S, Lin Y-C, Liu X-Y, et al.; Wafer-Scale and Deterministic Patterned Growth of Monolayer $MoS_2$ via Vapor–Liquid–Solid Method. *Nanoscale* 2019;**11**(34):16122-16129.

27. Shivayogimath A, Thomsen JD, Mackenzie DMA, et al.; A Universal Approach for the Synthesis of Two-Dimensional Binary Compounds. *Nat. Commun.* 2019;**10**(1):2957.

28. Ju M, Liang X, Liu J, et al.; Universal Substrate-Trapping Strategy to Grow Strictly Monolayer Transition Metal Dichalcogenides Crystals. *Chem. Mater.* 2017;**29**(14):6095-6103.

29. Lu Y, Chen T, Ryu G, et al.; Self-Limiting Growth of High-Quality 2D Monolayer $MoS_2$ by Direct Sulfurization Using Precursor-Soluble Substrates for Advanced Field-Effect Transistors and Photodetectors. ACS Appl. Nano Mater. 2018;**2**(1):369-378.

30. Prakash A, Singh M, Mishra R, et al.; Studies on Modified Borosilicate Glass for Enhancement of Solubility of Molybdenum. J. Non-Cryst. Solids. 2019;**510**:172-178.

31. Zhou J, Lin J, Huang X, et al.; A Library of Atomically Thin Metal Chalcogenides. *Nature* 2018;**556**(7701):355-359.

32. Dong J, Zhang L, Ding F; Kinetics of Graphene and 2D Materials Growth. *Adv. Mater.* 2019;**31**(9):e1801583.

33. Rao R, Islam AE, Singh S, et al.; Spectroscopic Evaluation of Charge-Transfer Doping and Strain in Graphene/$MoS_2$ Heterostructures. *Phys. Rev. B* 2019;**99**(19):195401.

34. Ahn GH, Amani M, Rasool H, et al.; Strain-Engineered Growth of Two-Dimensional Materials. *Nat. Commun.* 2017;**8**(1):608.

35. Lin Y, Ling X, Yu L, et al.; Dielectric Screening of Excitons and Trions in Single-Layer $MoS_2$. *Nano Lett.* 2014;**14**(10):5569-5576.

36. Liu B, Chen L, Liu G, et al.; High-Performance Chemical Sensing Using Schottky-Contacted Chemical Vapor Deposition Grown Mono layer $MoS_2$ Transistors. *ACS Nano*

2014;**8**(5):5304-5314.

37. Wang X, Gong Y, Shi G, et al.; Chemical Vapor Deposition Growth of Crystalline Mono layer MoSe$_2$. *Acs Nano* 2014;**8**(5):5125-5131.

38. Zhang R, Zhang Y, Zhang Q, et al.; Optical Visualization of Individual Ultralong Carbon Nanotubes by Chemical Vapour Deposition of Titanium Dioxide Nanoparticles. *Nat. Commun.* 2013;**4**:1727.

39. Splendiani A, Sun L, Zhang Y, et al.; Emerging Photoluminescence in Monolayer MoS$_2$. *Nano Lett.* 2010;**10**(4):1271-1275.

40. Tonndorf P, Schmidt R, Böttger P, et al.; Photoluminescence Emission and Raman Response of Monolayer MoS$_2$, MoSe$_2$, and WSe$_2$. *Optics Exp.* 2013;**21**(4):4908-4916.

41. Chen J, Zhao X, Tan SJR, et al.; Chemical Vapor Deposition of Large-Size Monolayer MoSe$_2$ Crystals on Molten Glass. *J. Am. Chem. Soc.* 2017;**139**(3):1073-1076.

42. Wang J, Luo X, Li S, et al.; Determination of Crystal Axes in Semimetallic T′-MoTe$_2$ by Polarized Raman Spectroscopy. *Adv. Funct. Mater.* 2017;**27**(14):1604799.

43. Gutiérrez HR, Perea-López N, Elías AL, et al.; Extraordinary Room-Temperature Photoluminescence in Triangular WS$_2$ Monolayers. *Nano Lett.* 2013;**13**(8):3447-3454.

44. Chen YF, Xi JY, Dumcenco DO, et al.; Tunable Band Gap Photoluminescence from Atomically Thin Transition-Metal Dichalcogenide Alloys. *Acs Nano* 2013;**7**(5):4610-4616.

45. Yun SJ, Duong DL, Doan M-H, et al.; Room-Temperature Ferromagnetism in Monolayer WSe$_2$ Semiconductor via Vanadium Dopant. *arXiv preprint arXiv:1806.06479 (2018)*.

FIGURE CAPTIONS

**Figure 1**. The DP growth of TMDCs in comparison with traditional CVD method. (a) Illustration of the DP growth process. The upper inset is an AFM image of one representative protrusion on the upper thin glass surface and the bottom inset is optical image of DP-grown $MoS_2$. (b-c) Comparisons between (b) DP growth and (c) traditional CVD growth of TMDCs in the nucleation process and the growth process.

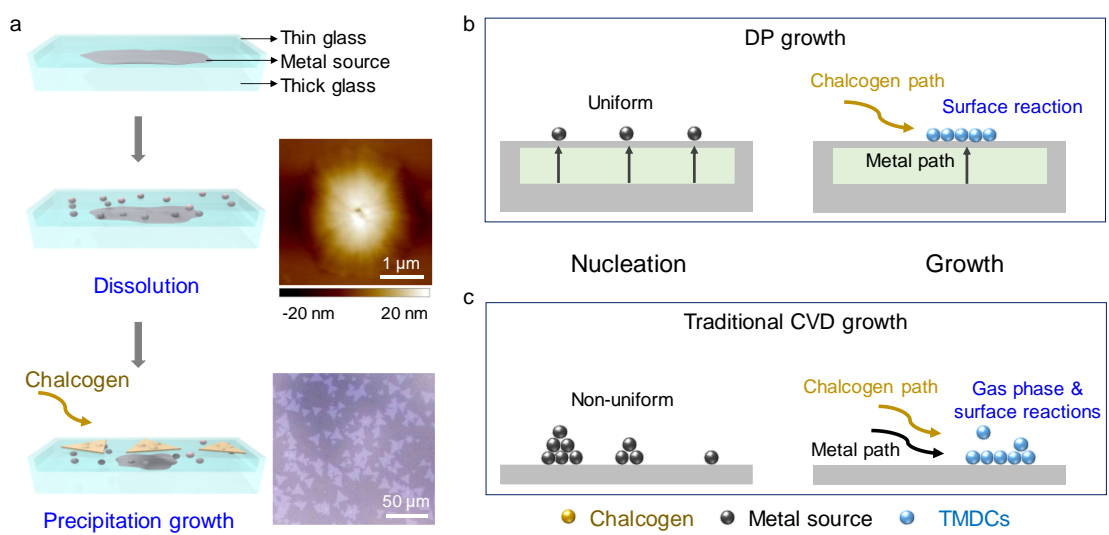

**Figure 2**. High uniformity of DP-grown monolayer MoS$_2$ flakes on a 2.5 x 1.0 cm$^2$ glass substrate. (a-b) Statistical analysis of (a) nucleation density and (b) average area of each MoS$_2$ flake from 140 images. Statistical analysis of (c) Raman peak positions and (d) PL intensity of 50 MoS$_2$ flakes randomly collected on a molten glass substrate. (e) An optical image and corresponding (f) PL map (the emission wavelength of 686.8 nm) and (g) Raman map (A$_{1g}$ peak at 403.1 cm$^{-1}$) of the as-grown MoS$_2$ flakes.

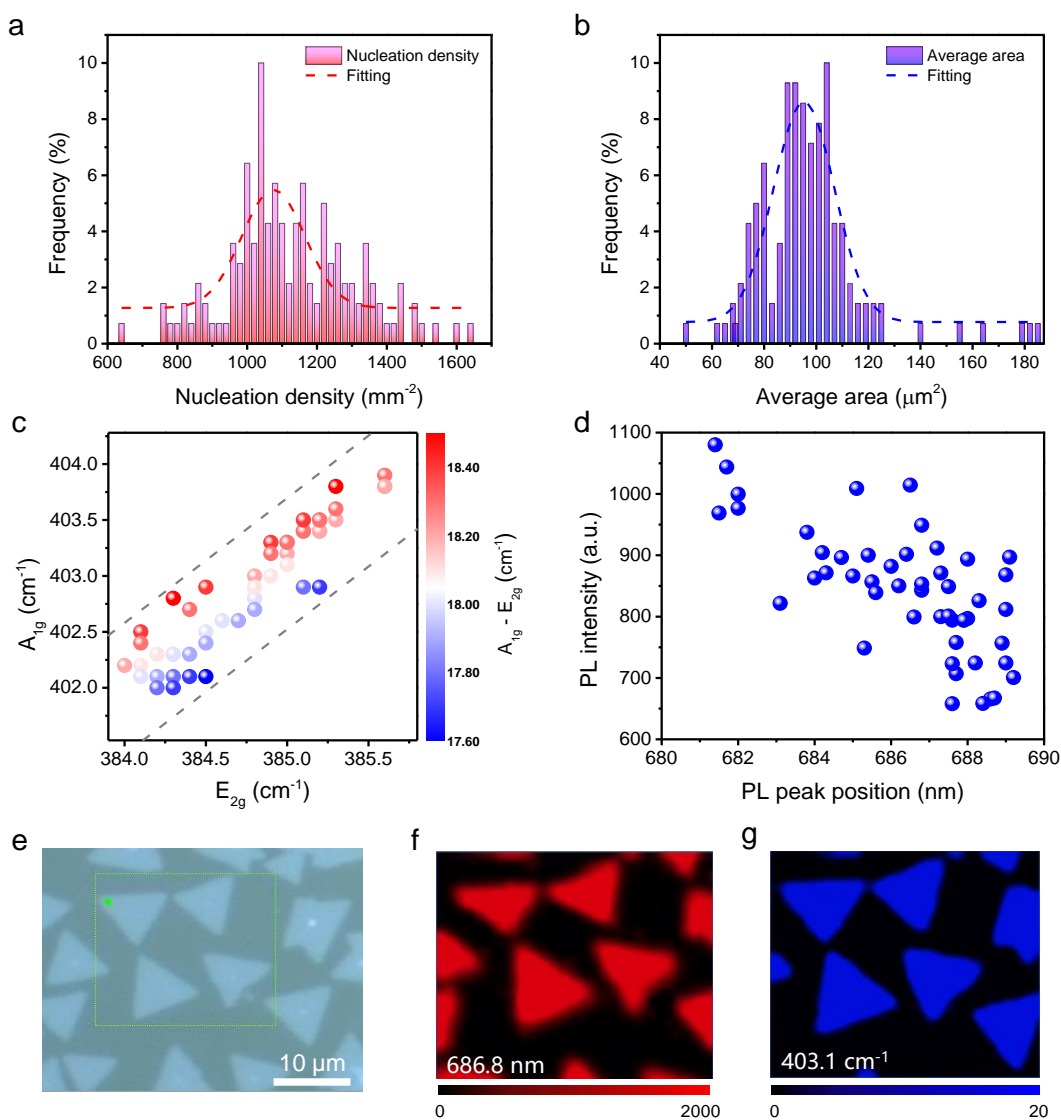

**Figure 3**. DP-grown MoS$_2$ with a clean surface and high quality. (a) AFM image of traditional CVD-grown MoS$_2$ with lots of particles absorbed at the edge and on the plane. (b) AFM image of DP-grown MoS$_2$ with a clean surface. (c) High resolution AFM image of DP-grown MoS$_2$ transferred onto a SiO$_2$/Si wafer. (d-e) Optical images of (d) traditional CVD-grown MoS$_2$ and (e) DP-grown MoS$_2$ after treatment with moisture TiCl$_4$ vapor. (f) PL intensity profile of as-grown monolayer MoS$_2$ and artificially stacked bilayer MoS$_2$. (g) STEM image of DP-grown MoS$_2$. Inset shows the corresponding FFT pattern. (h) Enlarged STEM image of the blue area denoted in (g). Inset shows the structural models of MoS$_2$. (i) Transfer curves of a field effect transistor made of the DP-grown monolayer MoS$_2$.

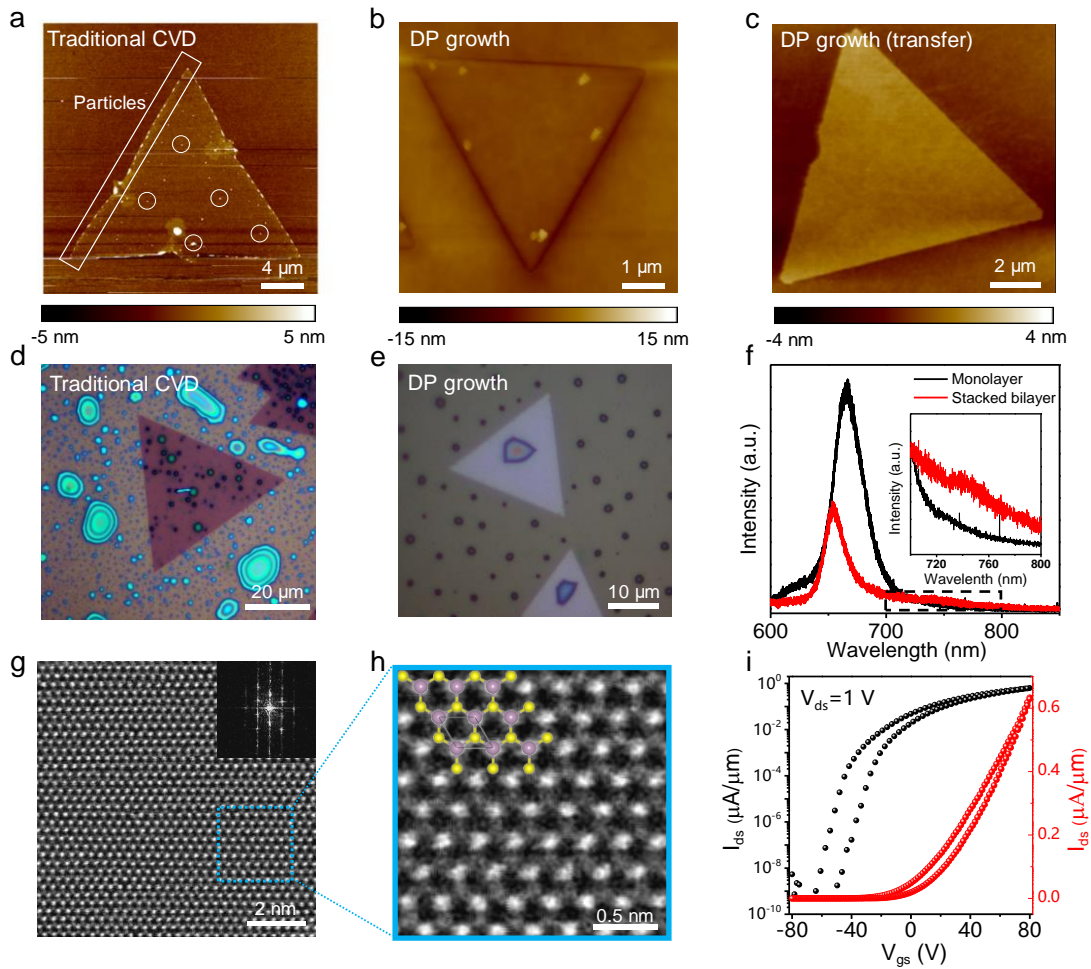

**Figure 4.** Universality of the DP method to grow different TMDCs. Optical images and corresponding Raman and PL spectra of DP-grown (a, b, c) MoSe$_2$ flakes, (d, e) 1T phase MoTe$_2$ flakes, (f, g, h) WS$_2$ flakes, and (i, k, k) Mo$_x$W$_{1-x}$S$_2$ alloy flakes.

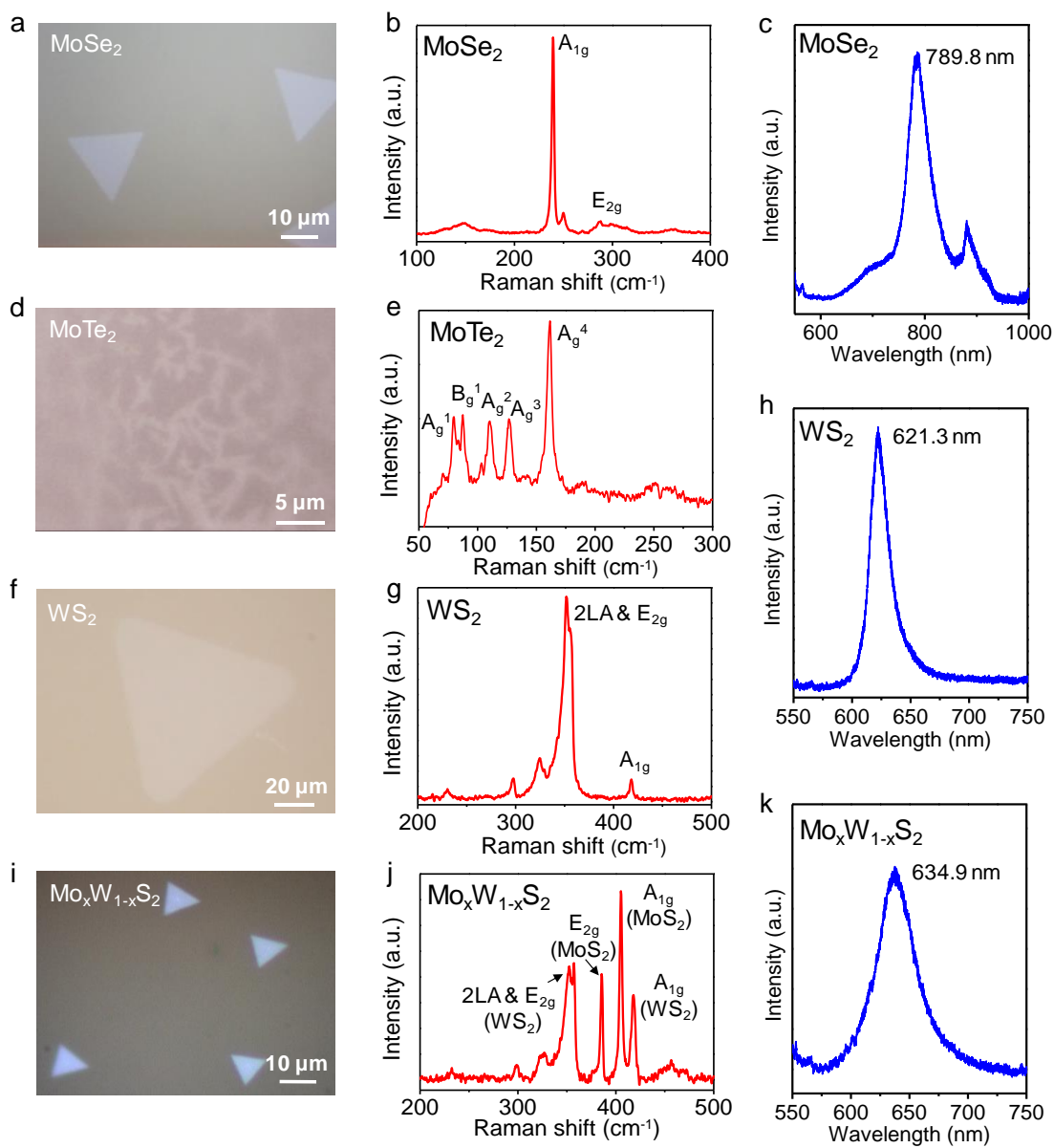